\documentclass[11pt]{article}
\usepackage{amsmath,amssymb,color,graphics,epsfig,cite}
\usepackage{graphicx,subfigure}
\usepackage{amsfonts}
\newcommand{\hoch}[1]{$\, ^{#1}$}

\newcommand{\be}{\begin{equation}}
	\newcommand{\ee}{\end{equation}}
\newcommand{\bea}{\setlength\arraycolsep{2pt} \begin{eqnarray}}
	\newcommand{\eea}{\end{eqnarray}}
\newcommand{\nn}{\nonumber}

\def\ft#1#2{{\textstyle{\frac{\scriptstyle #1}{\scriptstyle #2} } }}
\def\fft#1#2{{\frac{#1}{#2}}}

\def\0{{\sst{(0)}}}
\def\1{{\sst{(1)}}}
\def\2{{\sst{(2)}}}
\def\3{{\sst{(3)}}}
\def\4{{\sst{(4)}}}
\def\5{{\sst{(5)}}}
\def\6{{\sst{(6)}}}
\def\7{{\sst{(7)}}}
\def\8{{\sst{(8)}}}
\def\sst#1{{\scriptscriptstyle #1}}

\begin{document}
	
\begin{center}
		{\Large {\bf Black Hole Thermodynamics without Black Hole Solutions}}
		
		\vspace{20pt}
		
	Meng-Nan Yang\hoch{1}, Guan-Yi Lu\hoch{1} and H. L\"u\hoch{2,1}
		
		\vspace{10pt}

{\it \hoch{1}The International Joint Institute of Tianjin University, Fuzhou,\\ Tianjin University, Tianjin 300072, China}
\medskip

{\it \hoch{2}Center for Joint Quantum Studies, Department of Physics,\\
			School of Science, Tianjin University, Tianjin 300350, China }
%\medskip

%{\it \hoch{3}Peng Huanwu Center for Fundamental Theory, Hefei, Anhui 230026, China}
		
		\vspace{40pt}
		
		\underline{ABSTRACT}

	\end{center}

We consider the string-theory inspired Einstein-Maxwell-Maxwell-dilaton theory (EMMD) and show that we can derive the complete set of thermodynamic quantities of charged black holes, without having to solve for the black hole solutions. We argue that the technique can be applied more broadly to string theories, providing an accessible method for determining the thermodynamic properties of large classes of black holes for which exact solutions are typically unavailable.

\vfill {\footnotesize yangmengnan@tju.edu.cn\ \ \ lgy20020926@tju.edu.cn\ \ \ mrhonglu@gmail.com}
	
	\thispagestyle{empty}
	\pagebreak
	%\voffset=-40pt
	%\setcounter{page}{1}
	
	%\tableofcontents
	%\addtocontents{toc}{\protect\setcounter{tocdepth}{2}}
	
	%\newpage

\section{Introduction}
\label{sec:intro}

As one of the most important predictions of Einstein's theory of General Relativity, black holes have been a focus of both classical and quantum gravity. With Hawking's theoretical demonstration of the eponymous black hole radiation in the semiclassical approach \cite{Hawking:1974rv}, black hole dynamics \cite{Bardeen:1973gs} was promoted to become black hole thermodynamics, establishing a black hole as both a classical and quantum object. There is now a standard initial procedure in the construction and study of a black hole. For a given gravitational theory, typically in the Lagrangian formalism, we first derive the covariant equations of motion. We then propose an ansatz based on certain symmetries and substitute it into the equations of motion and obtain a set of consistent differential or partial differential equations. A black hole solution with required suitable asymptotic boundary conditions may emerge if there is an event horizon. We can then read off the black hole's thermodynamic information from both the asymptotic falloffs and the near-horizon geometry, including the mass, charges, temperature, entropy, chemical potentials, {\it etc.}, and verify its first law, which is guaranteed by the Wald formalism \cite{Wald:1993nt,Iyer:1994ys}. This is a rather deterministic and mechanical process in that when the theory is given, the outcome is completely specified and the procedure can be fully programmed.

We thus raise a natural question: can we determine the black hole thermodynamics directly from the theory, without going through this mechanical process? In this paper, we show, in certain theories inspired by strings, that we can indeed obtain the full black hole thermodynamic quantities without solving the field equations at all. For illustrative purposes, we shall focus on four spacetime dimensions. The low-energy effective theory of string is described by supergravity whose bosonic sector is composed of the metric and a set of $U(1)$ gauge fields, non-minimally coupled to scalar cosets that consist of dilatons and axions. One of the simplest but nontrivial examples is the STU supergravity model involving four $U(1)$ gauge fields, together with three $SL(2,\mathbb R)/SO(2)$ scalar cosets \cite{Duff:1995sm}. In this paper, we consider an STU model-inspired theory involving just one dilaton $\phi$ and two Maxwell fields $(A_1,A_2)$, with the Lagrangian
\be
{\cal L}= \sqrt{-g} \Big(R - \ft12 (\partial\phi)^2 - \ft14 e^{a_1 \phi} F_1^2 -
\ft14 e^{a_2\phi} F_2^2\Big),\label{emmd}
\ee
where $F_i = dA_i$. The electromagnetic dualities of both gauge fields imply that we can take the dilaton coupling constants  $a_1>0$ and $a_2<0$ without loss of generality. We shall consider spherically-symmetric and static black holes carrying electric charges of both Maxwell fields. When only one gauge field is turned on, the theory reduces to the well-known Einstein-Maxwell-dilaton (EMD) theory, and the GMGHS charged black hole solutions have been known for decades \cite{Gibbons:1987ps,Garfinkle:1990qj}. When both charges are turned on, exact solutions for general $(a_1,a_2)$ dilaton coupling constants are unknown, except for two special cases, namely $a_1=-a_2=-\sqrt3$ \cite{Gibbons:1985ac,Lu:2013uia} and $a_1a_2=-1$ \cite{Lu:2013eoa}. Note that when $a_1=-a_2$, the black hole with two electric charges is equivalent to a dyonic black hole of a single gauge field. In particular, the $(a_1,a_2)=(\sqrt3,-1/\sqrt3), (1,-1), (\sqrt3,-\sqrt3)$ examples can be embedded in the STU supergravity.

In this paper, we consider spherically-symmetric and static electrically-charged black holes of the EMMD theory \eqref{emmd}. Building on our earlier work \cite{Lu:2025eub}, we derive a complete set of the thermodynamic quantities without solving field equations. Our work was inspired by \cite{Cremonini:2023vwf}, where the mass/charge relation of the extremal dyonic black holes $(a_1=-a_2)$ was derived without solving the equations of motion. It is particularly advantageous to consider EMMD theory \eqref{emmd} to discuss the technique since it is nontrivial and yet there exist a sufficient number of exact solutions to test the results analytically. The paper is organised as follows. In Section 2, we list the assumptions used in our derivation. In Section 3, we present the results. In Section 4, we consider an application in testing thermodynamic bounds using the thermodynamic quantities we derive. We conclude the paper in Section 5.

\section{Assumptions and thermodynamic equations}
\label{sec:assume}

We restrict ourselves to the electrically-charged spherically-symmetric and static black holes that are asymptotic to Minkowski spacetime. The leading falloff at the asymptotic infinity for each field is given by
\be
g_{tt} \sim -1 + \fft{2M}{r} + \cdots\,,\qquad \phi \sim \fft{2\Sigma}{r} + \cdots\,,\qquad  A_i = \fft{4Q_i}{r} + \cdots\,,\qquad i=1,2\,.
\ee
In other words, the asymptotic spacetime is specified by the mass $M$, a scalar charge (hair parameter $\Sigma$) and two electric charges $Q_i$'s. We also assume that a black hole exists, with an event horizon which gives us $(T,S,\Phi_i)$, namely the temperature, entropy and two chemical potentials.

{\bf Long-range force law.} The long-range force between two identical black holes follows the inverse square law \cite{Cremonini:2022sxf}
\be
\lim_{r\rightarrow \infty} r^2\, \hbox{Force} = 4 Q_1^2 + 4 Q_2^2 - M^2 - \Sigma^2\,.
\ee
In the extremal limit, this long-range force vanishes, which was studied in \cite{Cremonini:2022sxf}. For general non-extremal black holes, where the force is nonzero, we introduce a new parameter $\mu$ to describe its strength \cite{Lu:2025eub}
\be
4 Q_1^2 + 4 Q_2^2 - M^2 - \Sigma^2=-\mu^2\,.\label{longrangeforce}
\ee
The advantage of introducing this new variable is that we can use $\mu$, instead of $M$, as a basic variable to describe a black hole.

{\bf Weak no-hair theorem conjecture.} We assume that the weak no-hair theorem conjecture \cite{Lu:2025eub} is valid such that, although the real scalar $\phi$ can be turned on in black hole construction, the scalar charge $\Sigma$ is not an independent parameter, but a function of the mass and charges. Such scalar hair is referred to as ``secondary hair'' in the literature. (See, e.g.~\cite{Herdeiro:2015waa}.) It then follows from \eqref{longrangeforce} that any thermodynamic quantity $X$ can be expressed in terms of the three basic variables $(Q_1,Q_2,\mu)$, namely $X=X(Q_1,Q_2,\mu)$. The dilaton constant shift symmetry implies that \cite{Lu:2025eub}
\be
\Sigma = \ft14 a_1 Q_1 \fft{\partial M}{\partial Q_1} + \ft14 a_1 Q_1 \fft{\partial M}{\partial Q_1} + \mu \fft{\partial \Sigma}{\partial \mu}\,.
\ee
This relation in the extremal case ($\mu=0$) was first proposed in \cite{Cremonini:2023vwf}, and it was generalized in \cite{Lu:2025eub} to the general non-extremal cases.

{\bf Black hole first law.} We then assume that the black hole thermodynamic first law is valid,
namely
\be
dM=TdS + \Phi_1 dQ_1 + \Phi_2 dQ_2\,.
\ee
This first law is guaranteed by the Wald formalism.

{\bf Asymptotic-horizon connection.} As was demonstrated in \cite{Lu:2025eub}, without having to solve the full field equations, the long-range net force strength $\mu$ is related to the horizon quantities by
\be
\mu=2 T S\,.
\ee
This property was also observed for black hole solutions in the EMD theory in \cite{Herdeiro:2018wub}, and the resulting relation \eqref{longrangeforce} was referred to as the quadratic Smarr relation.

{\bf Homogeneity.} Since the string-inspired EMMD theory has no dimensionful coupling constants, it follows that a thermodynamic quantity $X$ must be homogeneous under the scaling of the basic variables, namely
\be
X(\lambda Q_1, \lambda Q_2, \lambda \mu) = \lambda^n\, X(Q_1, Q_2,\mu)\,,
\ee
where $n$ is the dimension of the quantity $X$.

The mass/charge relation of extremal dyonic black holes in general EMD theory was obtained in \cite{Cremonini:2023vwf}. The relation was generalized to non-extremal black holes in general EMMD theory \cite{Lu:2025eub}. In the next section, we extend our work to deriving all the thermodynamic quantities including entropy and temperature, in terms of the three variables $(Q_1,Q_2,\mu)$, thereby deriving the black hole thermodynamics without solving the field equations.

\section{Results}

The homogeneity property implies that we can write the thermodynamic quantities as
\be
M=\mu f(x,y)\,,\quad \Sigma=\mu g(x,y)\,,\quad S=\mu^2 s(x,y)\,,\quad T=\fft{t(x,y)}{\mu}\,,\quad \Phi_i = \Phi_i(x,y)\,,
\ee
where $(x,y)=(Q_1,Q_2)/\mu$ are two dimensionless charge parameters. The properties discussed in Section \eqref{sec:assume} lead to the following set of algebraic and differential equations
\bea
&&1 + 4 x^2 + 4 y^2 - f^2 -  g^2 = 0\,,\qquad
(a_1 x \partial_x + a_2 y \partial_y) f -(x \partial_x + y \partial_y) g =0\label{master1}\,,\\
&&\ft12 (x \partial_{x} + y \partial_y) \log s =1 -  f +  (x \partial_x+y \partial_y) f\,,\label{master2}\\
&&\Phi_1 = \partial_x f - \ft12 \partial_x \log s\,,\qquad
\Phi_2 = \partial_y f - \ft12 \partial_y \log s\,,\qquad t = \fft1{2s}\,.\label{master3}
\eea
The two equations grouped in \eqref{master1} are self-contained and they determine the mass/charge relations \cite{Lu:2025eub}. We can then derive the entropy from \eqref{master2}, followed by the chemical potentials in \eqref{master3}. The temperature is simply $T=\mu/(2S)$. 

Following the strategy of \cite{Lu:2025eub}, we solve the equations of thermodynamic quantities progressively and use the simpler cases as boundary conditions for the more general cases. We first turn on only $Q_1$ by setting $Q_2=0$, and hence $y=0$. These equations can be straightforwardly solved and we have
\bea
M &=& \frac{\sqrt{4 \left(a_1^2+1\right) Q_1^2+\mu^2}+a_1^2\mu}{a_1^2+1}\,,\qquad
\Sigma=\frac{a_1 \left(\sqrt{4 \left(a_1^2+1\right) Q_1^2+\mu^2}-\mu\right)}{ 1+ a_1^2}\,,\nn\\
S &=& 4\pi \mu^2 \Big(\ft12 + \sqrt{\ft14 + (1+a_1^2) Q_1^2\mu^{-2}}\Big)^{\fft{2}{1+a_1^2}}\,,\qquad \Phi_1 = \fft{4Q_1}{\sqrt{4(1+a_1^2) Q_1^2 + \mu^2} + \mu}\,.
\eea
Together with $T=\mu/(2S)$, we obtain the full set of thermodynamic quantities of the GMGHS black holes, without solving for the black hole solutions. Although $\Sigma$ does not appear in the first law of black hole thermodynamics, it is nevertheless an important asymptotic parameter that dictates the long-range force. Note that the integration constant associated with entropy from \eqref{master2} is fixed by requiring that the entropy recovers that of the Schwarzschild black hole in the $Q_1=0$ limit.

For general two-charged solutions, we can use the above EMD single charge solutions for both $Q_2=0$ and $Q_1=0$ as boundary conditions to solve the differential equations \eqref{master1}, \eqref{master2} and \eqref{master3}. When $a_1 a_2=-1$, an exact solution exists and the mass and scalar charge were obtained, given by \cite{Lu:2013eoa}
\bea
M&=& \frac{\sqrt{4 \left(a_1^2+1\right) Q_1^2+\mu^2}}{a_1^2+1} + \frac{\sqrt{4 \left(a_1^{-2}+1\right) Q_2^2+\mu^2}}{a_1^{-2}+1}\,,\nn\\
\Sigma &=& \frac{a_1 \sqrt{4 \left(a_1^2+1\right) Q_1^2+\mu^2}}{2 \left(a_1^2+1\right)} -
\frac{a_1^{-1} \sqrt{4 \left(a_1^{-2}+1\right) Q_2^2+\mu}}{2 \left(a_1^{-2}+1\right)}\,.
\eea
We can then derive the rest from \eqref{master2} and \eqref{master3}, giving
\bea
S &=& \pi \left(\sqrt{4 \left(a_1^2+1\right) Q_1^2+\mu ^2}+\mu \right){}^{\frac{2}{a_1^2+1}} \left(\mu +\sqrt{\mu ^2+4 \left(a_1^{-2}+1\right) Q_2^2}\right){}^{\frac{2 a_1^2}{a_1^2+1}}\,,\nn\\
\Phi_1 &=& \frac{4 Q_1}{\mu+\sqrt{4 \left(a_1^2+1\right) Q_1^2+\mu ^2}}\,,\qquad
\Phi_2 = \frac{4Q_2}{\mu +\sqrt{\mu ^2+4 \left(a_1^{-2}+1\right) Q_2^2}}\,.
\eea
Another analytical example is when $a_1=-a_2=\sqrt3$, in which case, the field equations reduce to the $SL(3,\mathbb R)$ Toda equation \cite{Gibbons:1985ac,Lu:2013uia}. It is not convenient to express the thermodynamic quantities directly in terms of $(Q_1,Q_2,\mu)$. Instead, we can introduce a pair of charge parameters $(q_1,q_2)$, and the thermodynamic variables become \cite{Lu:2013uia}
\bea
M&=&\frac{1}{4} \left(4 \mu +q_1+q_2\right)\,,\qquad \Sigma=\frac{\sqrt3}{4} \left(q_1-q_2\right)\,,\qquad Q_i = \sqrt{\fft{q_i (q_i + 2\mu)(q_i+4\mu)}{16(q_1 + q_2 + 4\mu)}}\,,\nn\\
S &=& \frac{ \pi(q_1 + 4\mu)(q_2 + 4\mu) \sqrt{(q_1 + 2\mu)(q_2 + 2\mu)}}{
2 (q_1 + q_2 + 4\mu)}\,,\qquad
\Phi_i = \frac{\sqrt{q_i (q_1 + q_2 + 4\mu)}}{(q_i + 2\mu) (q_i + 2\mu)}\,.\label{sqrt3res}
\eea
It is easy to verify that these thermodynamic quantities indeed satisfy the equations (\ref{master1},\ref{master2},\ref{master3}). However, a priori, without strong mathematical proficiency, the equations (\ref{master1},\ref{master2},\ref{master3}) will not easily lend themselves to these simple analytical expressions in \eqref{sqrt3res}. Beyond these few examples of exact solutions, we need to solve (\ref{master1},\ref{master2},\ref{master3}) numerically for all the thermodynamic quantities, but these are simple equations and easy to solve.

\section{An application}

\begin{figure}[ht]
	\centering
	\includegraphics[width=0.45\linewidth]{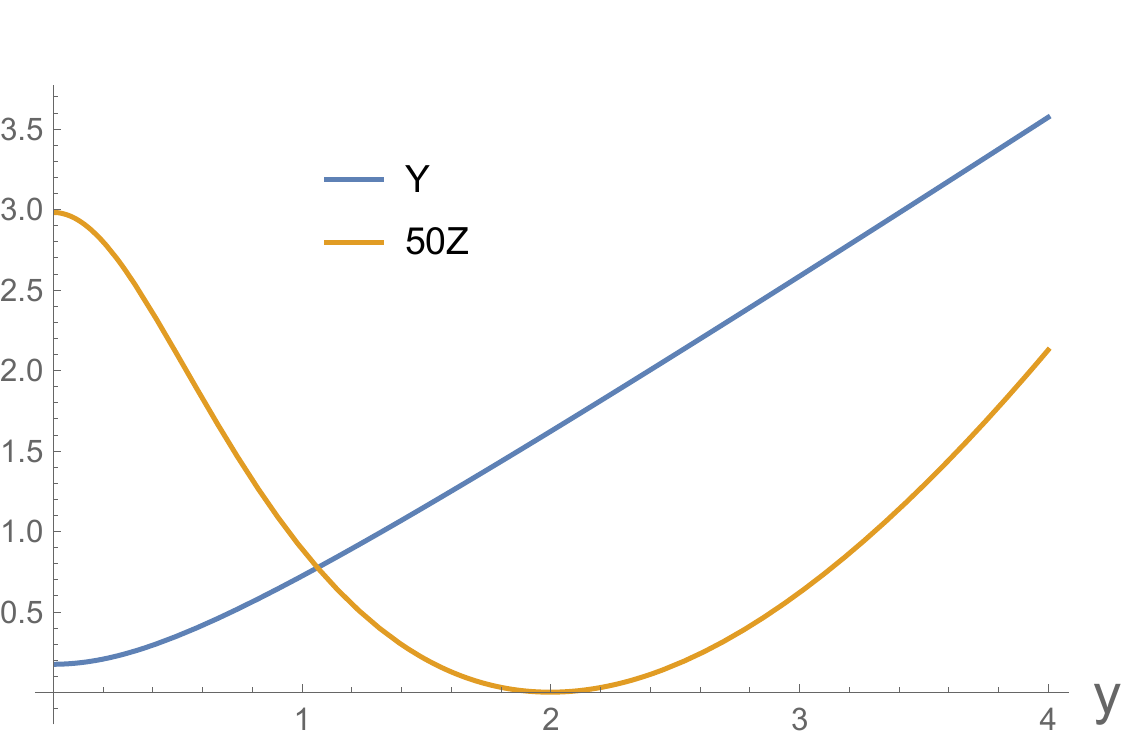} \includegraphics[width=0.45\linewidth]{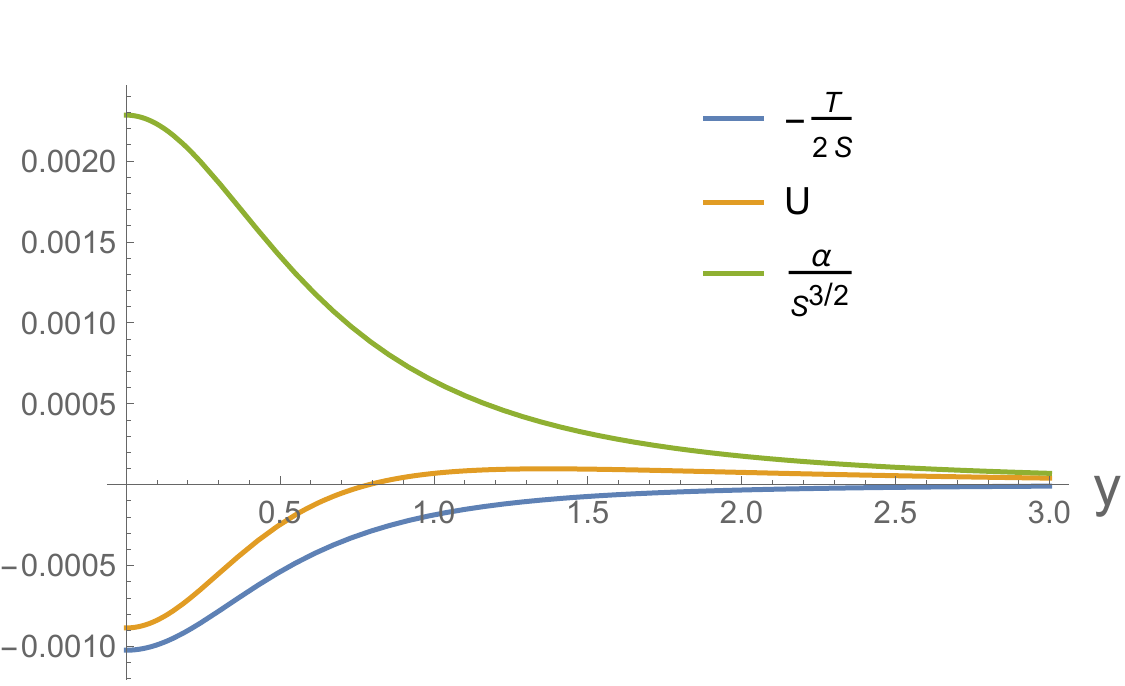}
	\caption{\small Both panels are associated with $(a_1,a_2)=(2,-1/8)$. We also set $x=1/2$ and $\mu=1$. In the left panel, we scaled the quantity $Z$ by 50 for clarity. We see that both $Y$ and $Z$ are nonnegative and $Z<Y$. In particular, when $y=2$, corresponding to the RN black hole, $Z$ vanished, as expected. In the right panel, we see that the line associated with $U$ is indeed sandwiched by its two bounds.}
	\label{fig:inequalities}
\end{figure}

We have shown that all the thermodynamic quantities of charged black holes in EMMD theories can be easily obtained in terms of $(Q_1,Q_2,\mu)$ variables. We now consider an application in testing the thermodynamic bounds of EMMD theories. The best-known bound is the Penrose entropy bound \cite{Bray:2003ns,Mars:2009cj}. For our static black holes, it states
\be
Y=M - \ft12 \sqrt{S/\pi}= \mu \Big(f(x,y) - \ft12\sqrt{s(x,y)/\pi}\Big)\ge 0\,.
\ee
A more stringent bound for static black holes involving further purely geometric quantities was proposed in \cite{Khodabakhshi:2022jot}, which states
\be
Z \equiv M + 2 T S - \sqrt{S/\pi} = \mu \Big(f(x,y) + 1 -\sqrt{s(x,y)/\pi}\Big)\ge 0\,.
\ee
The bound is saturated by both the Schwarzschild and RN black holes, and it is more stringent in the sense that $Z\le Y$. A sequence of entropy bounds based on the specific heat was recently proposed in \cite{Lu:2025iky}, which states
\be
-\ft12 T/S \le U \le \alpha S^{-3/2}\,,\qquad U \equiv \Big(\fft{\partial T}{\partial S}\Big)_{Q_i}\,,
\ee
where $\alpha = \fft{1}{4\sqrt\pi}$ for spherically-symmetric and static black holes. For general $(a_1,a_2)$, we can test these bounds numerically. Here we present an explicit example with $(a_1,a_2)=(2,-1/8)$, where the mass/charge relation was plotted in \cite{Lu:2025eub}. In particular, the RN black hole emerges when $y=4x$ in this case. It is easy to verify that all the bounds are satisfied. Instead of presenting a confirming 3D plot with $(x,y)$ variables that are not particularly instructive, we present some 2D plots with fixed $x=1/2$ in Fig.~\ref{fig:inequalities}. These plots confirming both the inequalities and the validity of our approach.

\section{Conclusions}

In the literature, some black hole thermodynamic properties can be obtained without having to solve the field equations. Notable examples include the first law from the Wald formalism \cite{Iyer:1994ys}. Another example is the entropy from the attractor mechanism \cite{Ferrara:1995ih,Sen:2005wa}. In this paper, we derived, for the first time, the complete set of thermodynamic quantities as functions of three basic variables $(Q_1,Q_2,\mu)$ of electrically charged black holes in EMMD theories, without solving the field equations.

It is natural to ask whether this technique can be applied to theories beyond EMMD. The assumptions discussed in Section \ref{sec:assume} are not so restrictive. However, if a theory contains additional coupling constants, such as a cosmological constant, the homogeneity condition no longer applies. Although this problem can be easily resolved by including the cosmological constant as a thermodynamic variable and introducing its conjugate as thermodynamic volume \cite{Kastor:2009wy,Cvetic:2010jb}, the concept of long-range force in AdS or dS spacetimes requires further modification. The inclusion of rotation can also be subtle. Nevertheless, it is reasonable to expect that leading-order supergravities that are low-energy effective theories of strings satisfy the criteria listed in Section \ref{sec:assume}, and we expect that our technique can be applied to derive all the thermodynamic quantities of charged black holes or more general $p$-branes.

\section*{Acknowledgement}

This work is supported in part by the National Natural Science Foundation of China (NSFC) grants No.~12375052 and No.~11935009, and also by the Tianjin University Self-Innovation Fund Extreme Basic Research Project Grant No.~2025XJ21-0007.


\begin{thebibliography}{99}

%\cite{Hawking:1974rv}
\bibitem{Hawking:1974rv}
S.W.~Hawking,
``Black hole explosions,''
Nature \textbf{248}, 30-31 (1974)
doi:10.1038/248030a0;
%3808 citations counted in INSPIRE as of 17 Jul 2022
``Particle creation by black holes,''
Commun. Math. Phys. \textbf{43}, 199-220 (1975)
[erratum: Commun. Math. Phys. \textbf{46}, 206 (1976)]
doi:10.1007/BF02345020
%9739 citations counted in INSPIRE as of 17 Jul 2022

%\cite{Bardeen:1973gs}
\bibitem{Bardeen:1973gs}
J.M.~Bardeen, B.~Carter and S.W.~Hawking,
``The four laws of black hole mechanics,''
Commun. Math. Phys. \textbf{31}, 161-170 (1973)
doi:10.1007/BF01645742
%2641 citations counted in INSPIRE as of 17 Jul 2022

\bibitem{Wald:1993nt}
R.M.~Wald,
``Black hole entropy is the Noether charge,''
Phys. Rev. D \textbf{48}, no.8, R3427-R3431 (1993), arXiv:gr-qc/9307038 [gr-qc].

\bibitem{Iyer:1994ys}
V.~Iyer and R.M.~Wald,
``Some properties of Noether charge and a proposal for dynamical black hole entropy,''
Phys. Rev. D \textbf{50}, 846-864 (1994), arXiv:gr-qc/9403028 [gr-qc].

%\cite{Duff:1995sm}
\bibitem{Duff:1995sm}
M.J.~Duff, J.T.~Liu and J.~Rahmfeld,
``Four-dimensional string-string-string triality,''
Nucl. Phys. B \textbf{459}, 125-159 (1996)
doi:10.1016/0550-3213(95)00555-2
[arXiv:hep-th/9508094 [hep-th]].
%303 citations counted in INSPIRE as of 13 Nov 2025

%\cite{Gibbons:1987ps}
\bibitem{Gibbons:1987ps}
G.W.~Gibbons and K.i.~Maeda,
``Black holes and membranes in higher dimensional theories with dilaton fields,''
Nucl. Phys. B \textbf{298}, 741-775 (1988)
doi:10.1016/0550-3213(88)90006-5
%1478 citations counted in INSPIRE as of 11 Aug 2025

%\cite{Garfinkle:1990qj}
\bibitem{Garfinkle:1990qj}
D.~Garfinkle, G.T.~Horowitz and A.~Strominger,
``Charged black holes in string theory,''
Phys. Rev. D \textbf{43}, 3140 (1991)
[erratum: Phys. Rev. D \textbf{45}, 3888 (1992)]
doi:10.1103/PhysRevD.43.3140
%1364 citations counted in INSPIRE as of 17 Aug 2025

%\cite{Gibbons:1985ac}
\bibitem{Gibbons:1985ac}
G.W.~Gibbons and D.L.~Wiltshire,
``Black holes in Kaluza-Klein theory,''
Annals Phys. \textbf{167}, 201-223 (1986)
[erratum: Annals Phys. \textbf{176}, 393 (1987)]
doi:10.1016/S0003-4916(86)80012-4
%320 citations counted in INSPIRE as of 11 Aug 2025

%\cite{Lu:2013uia}
\bibitem{Lu:2013uia}
H.~L{\"u} and W.~Yang,
``$SL(n,\mathbb R)$-Toda black holes,''
Class. Quant. Grav. \textbf{30}, 235021 (2013)
doi:10.1088/0264-9381/30/23/235021
[arXiv:1307.2305 [hep-th]].
%15 citations counted in INSPIRE as of 09 Dec 2025

%\cite{Lu:2013eoa}
\bibitem{Lu:2013eoa}
H.~L\"u,
``Charged dilatonic AdS black holes and magnetic AdS$_{D-2} \times R^{2}$ vacua,''
JHEP \textbf{09}, 112 (2013)
doi:10.1007/JHEP09(2013)112
[arXiv:1306.2386 [hep-th]].
%52 citations counted in INSPIRE as of 15 Aug 2025

%\cite{Lu:2025eub}
\bibitem{Lu:2025eub}
G.Y.~Lu, M.N.~Yang and H.~L\"u,
``Black hole mass/charge relation and weak no-hair theorem conjecture,''
JHEP \textbf{11}, 066 (2025)
doi:10.1007/JHEP11(2025)066
[arXiv:2508.14158 [hep-th]].
%2 citations counted in INSPIRE as of 04 Dec 2025

%\cite{Cremonini:2023vwf}
\bibitem{Cremonini:2023vwf}
S.~Cremonini, M.~Cveti\v c, C.N.~Pope and A.~Saha,
``Mass and force relations for Einstein-Maxwell-dilaton black holes,''
Phys. Rev. D \textbf{107}, no.12, 126023 (2023)
doi:10.1103/Phys RevD.107.126023
[arXiv:2304.04791 [hep-th]].
%2 citations counted in INSPIRE as of 11 Aug 2025

%\cite{Cremonini:2022sxf}
\bibitem{Cremonini:2022sxf}
S.~Cremonini, M.~Cveti\v c, C.N.~Pope and A.~Saha,
``Long-range forces between nonidentical black holes with non-BPS extremal limits,''
Phys. Rev. D \textbf{106}, no.8, 086007 (2022)
doi:10.1103/PhysRevD.106.086007
[arXiv:2207.00609 [hep-th]].
%2 citations counted in INSPIRE as of 30 Sep 2025

%\cite{Herdeiro:2015waa}
\bibitem{Herdeiro:2015waa}
C.A.R.~Herdeiro and E.~Radu,
``Asymptotically flat black holes with scalar hair: a review,''
Int. J. Mod. Phys. D \textbf{24}, no.09, 1542014 (2015)
doi:10.1142/S0218271815420146
[arXiv:1504.08209 [gr-qc]].
%722 citations counted in INSPIRE as of 31 Aug 2025

%\cite{Herdeiro:2018wub}
\bibitem{Herdeiro:2018wub}
C.A.R.~Herdeiro, E.~Radu, N.~Sanchis-Gual and J.A.~Font,
``Spontaneous scalarization of charged black holes,''
Phys. Rev. Lett. \textbf{121}, no.10, 101102 (2018)
doi:10.1103/Phys RevLett.121.101102
[arXiv:1806.05190 [gr-qc]].
%329 citations counted in INSPIRE as of 03 Dec 2025

%\cite{Bray:2003ns}
\bibitem{Bray:2003ns}
H.L.~Bray and P.T.~Chrusciel,
``The Penrose inequality,''
[arXiv:gr-qc/0312047 [gr-qc]].
%35 citations counted in INSPIRE as of 18 Jul 2025

%\cite{Mars:2009cj}
\bibitem{Mars:2009cj}
M.~Mars,
``Present status of the Penrose inequality,''
Class. Quant. Grav. \textbf{26}, 193001 (2009)
doi:10.1088/0264-9381/26/19/193001
[arXiv:0906.5566 [gr-qc]].
%141 citations counted in INSPIRE as of 18 Jul 2025

%\cite{Khodabakhshi:2022jot}
\bibitem{Khodabakhshi:2022jot}
H.~Khodabakhshi, H.~L{\"u} and R.~Q.~Yang,
``Tightening the Penrose inequality,''
Sci. China Phys. Mech. Astron. \textbf{65}, no.12, 120413 (2022)
doi:10.1007/s11433-022-2016-3
[arXiv:2207.08833 [gr-qc]].
%5 citations counted in INSPIRE as of 04 Dec 2025

%\cite{Lu:2025iky}
\bibitem{Lu:2025iky}
K.P.~Lu and H.~L\"u,
``Black hole entropy bounded by the specific heat,''
Phys. Rev. D \textbf{112}, no.8, 084030 (2025)
doi:10.1103/3hqz-45z5
[arXiv:2507.17812 [gr-qc]].
%0 citations counted in INSPIRE as of 04 Dec 2025

%\cite{Ferrara:1995ih}
\bibitem{Ferrara:1995ih}
S.~Ferrara, R.~Kallosh and A.~Strominger,
``$N=2$ extremal black holes,''
Phys. Rev. D \textbf{52}, R5412-R5416 (1995)
doi:10.1103/PhysRevD.52.R5412
[arXiv:hep-th/9508072 [hep-th]].
%999 citations counted in INSPIRE as of 04 Dec 2025

%\cite{Sen:2005wa}
\bibitem{Sen:2005wa}
A.~Sen,
``Black hole entropy function and the attractor mechanism in higher derivative gravity,''
JHEP \textbf{09}, 038 (2005)
doi:10.1088/1126-6708/2005/09/038
[arXiv:hep-th/0506177 [hep-th]].
%487 citations counted in INSPIRE as of 04 Dec 2025

%\cite{Kastor:2009wy}
\bibitem{Kastor:2009wy}
D.~Kastor, S.~Ray and J.~Traschen,
``Enthalpy and the mechanics of AdS black holes,''
Class. Quant. Grav. \textbf{26}, 195011 (2009)
doi:10.1088/0264-9381/26/19/195011
[arXiv: 0904.2765 [hep-th]].
%1397 citations counted in INSPIRE as of 04 Dec 2025

%\cite{Cvetic:2010jb}
\bibitem{Cvetic:2010jb}
M.~Cveti\v c, G.W.~Gibbons, D.~Kubiznak and C.N.~Pope,
``Black hole enthalpy and an entropy inequality for the thermodynamic volume,''
Phys. Rev. D \textbf{84}, 024037 (2011)
doi:10.1103/PhysRevD.84.024037
[arXiv:1012.2888 [hep-th]].
%713 citations counted in INSPIRE as of 04 Dec 2025

\end{thebibliography}
\end{document}